\documentclass{ws-mpla}

\begin{document}

\markboth{P. S. NEGI}
{Hydrostatic equilibrium of insular static spherically symmetric perfect fluid solutions}

%
\catchline{}{}{}{}{}
%

\author{P. S. NEGI\footnote{Ganga Sadan,
Sudarshan Hotel Compound, Nainital - 263 002, India.}}

\title{HYDROSTATIC EQUILIBRIUM OF INSULAR, STATIC, SPHERICALLY SYMMETRIC, PERFECT FLUID SOLUTIONS IN GENERAL RELATIVITY}

\address{Department of Physics, Kumaun University \\
Nainital, 263 002,
India\\
negi@upso.ernet.in}



\maketitle

\pub{Received (Day Month Year)}{Revised (Day Month Year)}

\begin{abstract}
   \abstract{
An analysis of insular solutions of Einstein's field equations for static, spherically symmetric, 
source mass, on the basis of exterior Schwarzschild solution is presented. Following the analysis, we demonstrate
that the {\em regular} solutions governed by a self-bound (that is, the surface density does not vanish together with
 pressure) equation of state (EOS) or density
variation can not exist in the state of hydrostatic equilibrium, because the source mass which belongs to them, does
not represent the `actual mass' appears in the exterior Schwarzschild solution.
The only configuration which could exist in this regard is governed by the homogeneous
density distribution (that is, the interior Schwarzschild solution). Other structures which naturally 
fulfill the requirement of the source mass, set up
by exterior Schwarzschild solution (and, therefore, can exist in hydrostatic equilibrium) are either 
governed by  gravitationally-bound regular solutions (that is, the surface density also vanishes
together with pressure), or self-bound  singular solutions
(that is, the pressure and density both become infinity at the centre).}

\end{abstract}

\ccode{PACS Nos.: 04.20.Jd; 04.40.Dg; 97.60.Jd.}

\section{Introduction}	

 The interior Schwarzschild solution 
(homogeneous density solution) of Einstein's field equations 
provides  two  very  important  features  towards  
obtaining {\em insular} (in the sense that the pressure vanishes at some finite boundary) configurations in  
hydrostatic  equilibrium,  compatible  
with general relativity, namely - (i) It gives an  absolute  upper  
limit on compactness ratio, $u (\equiv M/R$, mass  to  size  ratio  of  
the entire configuration in geometrized units) $\leq (4/9)$ for any
 static spherical configuration (corresponding to arbitrary density profile, provided the density decreases monotonically outwards from
 the centre) in hydrostatic equilibrium [1, 2], and (ii) For an assigned value of the  compactness ratio,  
$u$, and the radius, $R$,  the  minimum  central  pressure, $ P_0$,   corresponds   to   the  
homogeneous density solution (see,  e.g., ref.[2]). 
Recently, by using the property (ii) of homogeneous density sphere as mentioned above, we have 
obtained a `compatibility criterion for hydrostatic equilibrium' [3]. The important feature of this 
criterion is that it connects  the compactness ratio, $u$, of any static
configuration  with the 
corresponding ratio  of  central pressure  to   central   energy-density 
$\sigma [\equiv (P_0/E_0)$].
The criterion states that in order to have compatibility with the state of hydrostatic
equilibrium, for a given value of $\sigma$, the compactness ratio, $u$, of any 
configuration should always remain less than or equal  to  the compactness ratio
of the homogeneous  
density sphere for same $\sigma$ [3].
 
      Various insular exact solutions [4, 5; and references therein] and equations of state (EOSs) for  static  and  spherically  
symmetric mass are available in the literature which can be divided, in general, into two categories: 

\medskip

1. The exact solutions and EOSs corresponding
to the {\em regular} density variation [regular in the sense of positive finite density
at the origin (i. e., the metric coefficient, $e^{\lambda} = 1$ at $r = 0$, defined later)
which decreases monotonically outwards], such that the density at the surface vanishes together
with pressure (and so called the gravitationally-bound regular structures),
and

\medskip
 
2. The exact solutions and EOSs
corresponding to the density variation such that the density does not terminate
together with pressure at the surface of the configuration (and so called the self-bound
structures). These structures can be divided further into two sub-categories:

\smallskip

(a) The self-bound structures with finite central densities, generally called, the self-bound
regular structures\footnote[1]{There exists many self-bound exact solutions in the literature which also belong
to finite central densities, but what is required to the present context is ``the monotonic decrease of density
outwards from the centre'', and these solutions, e.g. - Goldman I solution [6] (called Gold I in Delgatey-Lake 
classification [5]), Stewart's solution [7], and Durgapal-Pande-Phuloria I
solution [8] (called D-P-P I in [5]) etc. do not fulfill this criterion [5]. Apparently, such solutions are irrelevant to the context of the present
study.}

\smallskip

(b) The self-bound structures with infinite central densities (i. e., $e^{\lambda} \neq 1$ at $r = 0$). 
We can call them, 
the self-bound singular structures (singular in the sense that pressure and density
both become infinity at the centre).

The exact solutions in the first category include Tolman's type VII solution with vanishing surface density
[9, 10, 11; and references therein], and
Buchdahl's ``gaseous" model\footnote[2]{This model represents the gravitationally-bound regular structure
for $u$ values $\leq 0.20$ [12].}[12], whereas the EOSs in this category include the well known
polytropic EOSs [13, 14]. The exact solutions in the second, sub-category (a) include 
Tolman's type IV
solution [9], the solution independently obtained by Adler [15], Kuchowicz [16],
and Adams and Cohen [17],
and Durgapal and Fuloria solution [18] etc. The well known example of EOS in this category is 
characterized
 by the stiffest EOS (see, e. g., ref. [19], [20])
$(dP/dE) = 1$ (in  geometrized  units). Haensel and Zdunik [20] have shown that the  only  EOS  which  
can describe a sub-millisecond pulsar and the static mass of $1.442  M_\odot $  
simultaneously, corresponds to the said stiffest EOS, however, they 
emphasized that this EOS represents an `abnormal' state of matter 
in the sense that pressure vanishes at densities of the order  of 
nuclear density or even higher [21]. The exact solutions in the second, sub-category
(b) include Tolman's type V and VI solutions [9], and the well known
example of EOS in this category is represented by the EOS corresponding to a `Fermi gas' having
infinite values of pressure and density at the centre [22].

An examination of the `compatibility criterion' on some well  
known exact solutions and EOSs indicated  that  
this criterion, in fact, is fulfilled only by those structures which come under
the category 1 and 2(b) mentioned above, that is, (i) the gravitationally-bound regular,
and, (ii) the self-bound singular structures.
 On the other hand, it is seen that 
the EOSs and analytic  solutions,  corresponding the self-bound regular state of matter 
[which are mentioned above under the category 2(a)],
in fact, do  not fulfill this criterion [3]. We have shown this 
inconsistency particularly for the EOS, $(dP/dE) = 1$
 (as it represents the most successful EOS to 
obtain the various extreme characteristics of neutron stars mentioned above),
and the analytic solution put forward by Durgapal and Fuloria [18] which fulfills
various properties for physically realistic structure [23].

     The  reason behind non-fulfillment of the `compatibility criterion'
by various self-bound regular EOSs and exact solutions could be resolved, if we carefully analyze
the `specific property' of the total mass $`M'$ appears in the exterior  Schwarzschild solution.
It immediately follows from this analysis that unlike gravitationally-bound regular structures,
and self-bound singular solutions, the `actual' total mass $`M'$, in fact,
 can not be attained by the configurations corresponding to a regular self-bound state of matter.
 This is demonstrated by considering a `generalized density distribution' for the source
  mass $`M'$ and verified further on the basis of a `class' of well known exact solutions, generated by what is called,
  `the algorithmic construction of all static spherically symmetric perfect fluid solutions of Einstein's
  equations' [24-26].



\section{Field Equations and TOV Equations}
For  a  spherically  symmetric  and static line element
\begin{equation} ds^2  = e^\nu dt^2  - e^\lambda dr^2  - r^2 d\theta^2 - r^2\sin^2 \theta d\phi^2,\end{equation}
[$\nu$ and $\lambda$ are functions of $r$ alone] recalling that we are using geometrized units, the  
field  equations yield in the following form
\begin{eqnarray} 
  8\pi T^0_{0} & = & 8\pi E = e^{-\lambda} [(\lambda'/r) - (1/r^2 )] + 1/r^2, \\
- 8\pi T^1_{1} & = & 8\pi P = e^{-\lambda} [(\nu'/r) + (1/r^2)] - 1/r^2, \\
- 8\pi T^2_{2} & = & - 8\pi T^3_{3} = 8\pi P = e^{-\lambda} [(\nu''/2) \nonumber  \\
               &   & +(\nu'^2/4)-(\nu'\lambda'/4)+(\nu'- \lambda')/2r]. 
\end{eqnarray}

where the primes represent differentiation with respect to $r$.  $P$ 
and $E$ represent, respectively,  the  pressure  and  energy-density 
inside the perfect fluid sphere  related  with  the  non-vanishing 
components of the energy-momentum tensor, $T_i^j$ =  0,  1,  2, 
and 3 respectively.
     Eqs. (2) - (4) represent second-order,  coupled  differential 
equations  which  can  be  written in  the form of first-order,  coupled 
differential equations, namely,  TOV equations [9, 22]
governing hydrostatic equilibrium in general relativity
\begin{equation}
P'  =  -(P + E)[4\pi Pr^3  + m(r)]/r(r - 2m(r)),                  
\end{equation}
\begin{equation}
\nu'  =  - 2P'/(P + E),                                          
\end{equation}
and
\begin{equation}
m'(r)  =  4\pi Er^2 ,                                               
\end{equation}
where prime denotes differentiation with respect to $r$, and $m(r)$
is the mass-energy contained within the radius $`r'$, that is

\begin{equation}
m(r)  =  \int_{0}^{r} 4\pi Er^2 dr.                                            
\end{equation}   

The equation connecting metric parameter $\lambda$ with $m(r)$ is given by

\begin{equation}
e^{-\lambda}    =  1  -  [2m(r)/r] = 1 - (8\pi/r)\int_{0}^{r} Er^2 dr.                                   
\end{equation}

The three field equations (or TOV equations) mentioned 
above, involve four variables, namely, $P, E, \nu,$ and  $\lambda$.  Thus,  in 
order to obtain a solution of these equations, one more equation is needed which may 
be assumed as a relation between $P$, and $E$ (EOS), or can be  regarded  as 
an algebraic relation connecting one of the  four  variables  with 
the radial coordinate $r$ (or an algebraic relation between the parameters)].
For obtaining an exact solution, the later approach is employed.  

Notice  that Eq. (9) yields the metric coefficient $e^{\lambda}$ for the assumed
energy-density, $E$, as a function of radial distance $`r$'. Once the metric coefficient
$e^{\lambda}$ or mass $m(r)$ is defined for assumed  energy-density by using Eqs. (9) or (8), 
the pressure, $P$, and the metric coefficient, $e^{\nu}$,
can be obtained by solving Eqs. (5) and (6) respectively which yield
two constants of integration. These constants should be obtained from the following boundary 
conditions, in order to have a proper solution of the field equations:

\section{Boundary Conditions: Hydrostatic Equilibrium for the Mass Distribution 
}

 In order to maintain hydrostatic  equilibrium  throughout  the 
configuration, the pressure must vanish  at  the  surface  of  the 
configuration, that is
\begin{equation}
P  =  P(r = R)  = P(R) =  0,                                     
\end{equation}
where $`R'$ is the radius of the configuration.

The consequence of Eq. (10) ensures  the  continuity  of $\nu'$, and, therefore, that of the
metric parameter $e^\nu$, belonging to the interior solution with  the  
corresponding expression for  well  known  exterior  Schwarzschild  
solution at the surface  of  the  fluid  configuration,  that  is:  
$e^{\nu(r = R)} = 1 - (2M/R)$ [where $`M' = m(R)$ is the total mass  of  the  
configuration].  However, the exterior Schwarzschild solution guarantees that
$e^{\nu(r = R)}$ = $e^{-\lambda(r = R)}$, at the surface  of  the  
configuration irrespective of the condition that the surface density, $E(r = R) = E(R)$,
is vanishing with pressure or not, that is
\begin{equation}
 E(R) = 0,                                     
\end{equation}
together with Eq. (10), or
\begin{equation}
 E(R) \neq 0,                                     
\end{equation}

So that one could assure the well known relation

\begin{equation}
e^{\nu(R)}  =  e^{-\lambda(R)}  =  1  - (2M/R)  =  (1 - 2u), 
\end{equation}
at the surface of the configuration, for both of the cases, namely - (i) the surface density vanishes together
with pressure, and (ii) the surface density does not vanish together with pressure. The total mass $M$ which appears
in Eq. (13) is defined as [Eq. (8)] 
 
\begin{equation}
M = m(R)  =  \int_{0}^{R} 4\pi Er^2 dr.                                            
\end{equation}
But, the most important point which should be remembered here is the `specific property'
of the total mass $`M = m(R)'$ follows directly from the well known property of 
the exterior Schwarzschild solution, namely - it depends only upon the total mass
$`M'$ and not upon the `type' of the density variation considered inside the 
radius $`R'$ of this mass generating sphere. That is, the total mass $`M'$ should
also bear this well known property of the exterior Schwarzschild solution, 
what we call it, the `type independence' property of the mass $`M'$, and it may be defined
in this manner: ``The dependence of mass $`M'$ upon the parameter(s) describing a particular
type of the 
density distribution considered inside the mass generating sphere (e. g., for an assumed value of the
compactness ratio, $u$, the mass, $`M'$, will depend only upon the radius, $R$, which may either
depend upon the surface density, or upon the central density, or upon both of them) should exist in 
such a manner 
that from an exterior observer's point of view, we could not diagnose that what `type' of
density variation belongs to this mass''.

This point can be illustrated in the following manner: the total mass $`M'$ which appears
in the exterior Schwarzschild solution is called the coordinate mass - the mass
measured by some external observer, and from this observer's point of view, if we measure
a sphere of mass $`M'$, we can not know (by any means) that how the matter is distributed
from centre to the surface of this sphere. It means, if we `measure' $M$ by the use of the 
non-vanishing surface density [we can calculate the (coordinate) radius $`R'$ by assigning a
suitable value to the surface density in the respective expression and by using the relation,
$M = uR$, the mass $`M'$ can be worked out for an assumed value of $u$], we can not measure
it, by any means, by the use of the central density (in order to keep the `type independence'
property of the mass intact), and this is possible only when there 
exist no relation connecting $`M'$ and the central density. Or, in other words, $`M'$ should
be independent of the central density (meaning thereby that the surface density should be independent
of the central density). On the other hand, if we measure $`M'$ by using the expression for 
central density [by assigning a suitable value to the central density in the relevant expression,
we can calculate the radius $`R'$ of the configuration, and by using the relation, $M = uR$, we can
work out the mass $`M'$ for an assumed value of $u$], we can not be able to measure $`M'$ by using
the expression for surface density (in order to keep the `type independence'
property of the mass intact). That is, there should exist no relation connecting the total mass
with surface density. It follows, therefore, that the central density should be independent of the 
surface density.

From the above explanation of `type independence' property of mass $`M'$, it is evident that
{\em the `actual' total mass $`M'$ which appears in the exterior Schwarzschild solution should either
depend upon the surface density, or depend upon the central density of the configuration, and in any
case, not upon both of them}. However, the dependence of mass $`M'$ upon both of the densities (surface,
as well as central) is a common feature observed among all self-bound regular structures, composed of
a single EOS or an analytic solution (which is the violation of the `type independence' property of mass
$`M'$), such structures, therefore, do not correspond to the `actual'
total mass $`M'$ required by the exterior Schwarzschild solution for the fulfillment of the boundary
conditions at the surface of the structure. This also explains the reason behind non-fulfillment of
the `compatibility criterion' by them. It is interesting to note here that there could exist only one
solution in this regard for which the mass $`M'$ depends upon both, but the same value of surface and
centre density, and for regular density distribution the structure would be governed by the homogeneous (constant)
density throughout the configuration.

For gravitationally bound regular structures, the requirement `type independence' of the mass  is obviously
fulfilled because the mass $`M'$ depends only upon the central density (surface density is always zero
for these structures). Furthermore, this demand of `type independence' is also satisfied by the self-bound 
singular solutions, because such structures correspond to an infinite value of central density, and consequently, 
the mass
$`M'$ depends only upon surface density. Both types of these structures are also found to be consistent
with the `compatibility criterion' as mentioned earlier.

The above discussion regarding various types of structures is true for any single equation of state or 
analytic solution comprises the whole configuration. At this place, we are not claiming that
the construction of a self-bound regular structure in impossible. It is quite
possible, provided we consider a two-density structure, such that the mass
$`M'$ always turns out to be independent of the central density, and the property `type independence'
of the mass $`M'$ is satisfied. Examples
of such two-density models are also available in the literature (see, e.g.,
ref. [27]), but in the different context. However, it should be noted here that {\em the fulfillment
of `type independence' condition by the mass $`M'$ for any two-density model will
represent only a necessary condition for hydrostatic equilibrium, unless the
`compatibility criterion' [3] is satisfied by them, which
also assure a sufficient and necessary condition for any structure
in hydrostatic equilibrium} (however, this
issue is discussed in detail elsewhere [28]), we restrict
ourself to the present context. In the following sections (4, and 5), explicit examples are
given to show that a single density variation comprising a self-bound regular
structure can not correspond to the actual mass $`M'$ required for the
hydrostatic equilibrium.

\section{The Actual Mass $M$ and the Generalized Density Distribution 
}

Now, consider the case of regular self-bound structures:
The  most smooth  possible  variation  of  
density inside  any regular configuration  can not be other than the
constant  (homogeneous)  density,  whereas  the  fastest  possible  
variation of density is well known and represented by  the  inverse  square  density  
variation  $[E \propto (1/r^{2})]$. It follows,  
therefore,  that  any possible regular self-bound configuration, characterized by an EOS
or, density as a function of radial co-ordinate, can be generalized in the following form \footnote[3]{One may
use some other form of the equation to generalize the self-bound regular structures considered
here, for example, we may consider a particular form of the `source function' [see, next section] which could
`generate' a `class' of solutions of the type considered here [category 2(a) of the present study] in the 
technique called `the algorithm for constructing all static spherically 
symmetric perfect fluid solutions' [24-26]. However, the conclusions drawn on this basis will remain unaltered, 
because the key point is that
for any self-bound regular configuration governed by a {\em single} EOS or density distribution, the surface
density can not be independent of the centre density. This is what we have demonstrated in the present, and also 
in the next section.}
 \begin{equation}
E(r) = C/{(a + r)}^b                               
 \end{equation}
where $C$ is the constant of proportionality, $a$ is a positive arbitrary constant
to make the density positive finite at the centre, and the constant $b$ is allowed to take
any value in the interval $0  \leq  b  \leq  2$.  Eq.  (15)  
represents a self-bound regular density  distribution,  the  density  at  the  
centre is positive finite, decreases monotonically from centre  to  
the outer region, and it  would  remain  finite  non-zero  if  one  
assumes that pressure vanishes at some finite radius. Thus, by using these central
and surface conditions in Eq. (15), we get
\begin{equation}
C = E_R{(a + R)}^b,
\end{equation}
and
\begin{equation}
C = E_0{a}^b
\end{equation}
where $E_0$ is
the central density, $R$ is the radius of the configuration, and  $E_R$ is the surface density 
at which pressure vanishes.

Substituting Eq. (15) into Eq. (8), we get the mass contained within the radial
co-ordinate $`r'$ for the assigned density variation as
\begin{equation}
m(r) = 4\pi C \int_{0}^{r}\frac{r^2dr}{{(a + r)}^b} = 4\pi CrI_r,
\end{equation}
where $I_r$ is given by
\begin{equation}
I_r = \frac {r}{(3 - b)(a + r)^{(b - 1)}} - \frac {2a}{r(3 - b)} \int_{0}^{r}\frac {rdr}{(a + r)^b}.
\end{equation}
The substitution of Eq. (18) into Eq. (9), yields the metric parameter, $e^\lambda$, for
the assigned density variation [Eq. (15)] as
\begin{equation}
e^{-\lambda} = 1 - 8\pi CI_r,
\end{equation}
where $C$ is defined by Eqs. (16) and (17) respectively, and $I_r$ is given by Eq. (19).
For Eq. (15), the total mass contained inside radius $`R'$ [Eq. (14)] is given by
\begin{equation}
M = 4\pi CRI_R.
\end{equation}
Eq. (20) gives $e^\lambda$ at the surface of the configuration as
\begin{equation}
e^{-\lambda(R)} = 1 - 8\pi CI_R
\end{equation}
where $I_R$ is given by
\begin{equation}
I_R = \frac {R}{(3 - b)(a + R)^{(b - 1)}} - \frac {2a}{R(3 - b)} \int_{0}^{R}\frac {rdr}{(a + r)^b}.
\end{equation}
The condition of regularity requires that
\begin{equation}
E_R \leq E_0.
\end{equation}
Substituting the values of $E_R$ and $E_0$ from Eqs. (16) and (17) into Eq. (24), we get
\begin{equation}
a^b \leq (a + R)^b.
\end{equation}

(A)The condition of equality in Eq (25) [that is, $a^b = (a + R)^b$] requires that
\begin{equation}
b = 0.
\end{equation}
For this value of $b(=0)$, Eqs. (16) and (17) give the relation
\begin{equation}
E_R = E_0 = C ({\rm constant}) = <E> ({\rm say}).
\end{equation}
The substitution of Eqs. (26) and (27) into Eq. (20) gives the metric parameter
$e^\lambda$ as
\begin{equation}
e^{-\lambda} = 1 - (8\pi/3) <E>r^2,
\end{equation}
which reduces at the surface of the configuration naturally to the value 
prescribed by exterior Schwarzschild solution
\begin{equation}
e^{-\lambda(R)} = 1 - (8\pi/3) <E>R^2 = 1 - (2M/R).
\end{equation}
The constant $C$ in this case is defined by both, but the same values of
surface and central densities [Eq. (27)] which represents the (constant) density
throughout the configuration for regular solution, given by [see, Eq. (14)] 
\begin{equation}
 <E> = 3M/4\pi R^3.
\end{equation}
Hence, the mass $`M'$  fulfills the property of `type independence' in this case
as discussed in the last section, therefore, represents the `actual mass'
which appears in the exterior Schwarzschild solution.

(B) Now, the condition of
inequality in Eq (25) [that is, $a^b < (a + R)^b$] requires that
\begin{equation} 0 < b \leq 2. \end{equation}
For these values of $b$ [Eq. (31)], we get from Eqs. (22), the value of
$e^\lambda$ at the surface of the configuration as
\begin{equation}
e^{-\lambda(R)} = 1 - \frac {8\pi C}{(3 - b)} \Bigl [\frac {R}{(a +
R)^{(b - 1)}} - \frac {2a}{R} \int_{0}^{R}\frac {rdr}{(a + r)^b} \Bigr].
\end{equation}

The right-hand side of Eq. (32) can not attain the normal value, $1 - (2M/R)$, prescribed by 
the exterior Schwarzschild
solution, because the constant $C$ in this case has been defined by both of the densities (surface, as
well as central) through Eqs. (16) and (17) respectively, therefore, it can not be eliminated trivially
from Eq. (32) 
by just using Eq. (21) [in the manner, $CI_R = (M/4\pi R)$](this common practice has been adopted by various authors in this regard] in order
to keep the `type independent' property of the mass $`M'$ intact. Or, in other words, the total mass
of a self-bound regular configuration given by Eq. (21) does not represent the `actual mass' $M$ appears
in the exterior Schwarzschild solution and consequently, the boundary conditions at the surface of such
configuration are not satisfied.

If, however, we set $a = 0$ in Eq. (15), we get from Eqs. (16) and (17), the relations
\begin{equation}
C = E_R R^b
\end{equation}
and,
\begin{equation}
E_0 = \infty.
\end{equation}

By substituting the value of $C$ from Eq.(33) into Eq.(32), we get $e^\lambda$ at the surface of 
the structure as 
\begin{equation}
e^{-\lambda(R)} = 1 - \frac{8\pi E_RR^2}{(3 - b)}
\end{equation}
or,
\begin{equation}
e^{-\lambda(R)} = 1 - (8\pi/3) <E>R^2 = 1 - (2M/R)
\end{equation}
which is the normal value required by exterior Schwarzschild solution at the surface.
In this equation, $<E>$ represents the `average density' of the structure given by
\begin{equation}
 <E> = \frac{3E_R}{(3 - b)} = \frac{3M}{4\pi R^3}.
\end{equation}

Note that for a calculated value of $`R'$, the mass $`M'$ in this case depends only upon the surface density
and not upon the central density (which is always infinite), thus the property `type independence' of
the mass is satisfied in this case and the structures will be compatible with the state of hydrostatic
equilibrium. 

\section{The Actual Mass $M$ and a `Class' of Exact Solutions Generated in the Algorithmic Construction
 }
 
The results of the analysis stated under the section 3 of the present study may be discussed in the context of 
the recent study, generally called `the algorithm for the construction of all static 
perfect fluid solutions in general relativity' [24-26], because it
 does not require any specific knowledge of the EOS (instead, it results as a byproduct of the algorithm).
The basis of such type of study is the well known fact that the 
pressure isotropy condition in perfect fluid configurations places a single differential constraint on the 
metric components of the geometry of the perfect fluid. It follows, therefore, that the class of metrics 
representing a perfect fluid geometry should be specified by a single arbitrary monotone function called 
the `generating function'. Rahman and Visser [24] and Lake [25] have shown that an explicit 
closed-form (algebraic-integro-differential) solution of the pressure isotropy condition in terms of this 
function, in fact, exists. Rahman and Visser [24] have presented their algorithm in 
`isotropic coordinates', because of the usefulness of this system, whereas Lake [25] has considered the 
`curvature coordinates' (which offer a direct physical interpretation of the `generating function') and also 
transformed it into `isotropic coordinates'. Martin and Visser [26] have presented a variant of the Lake's 
algorithm [25] in terms of the `average density' and the locally measured `acceleration due to gravity', that is, 
the variables with a clear physical meaning.

Although, the details towards choosing the `generating functions', corresponding to `physically reasonable' solutions 
[5, 25]\footnote[4]{The `physically reasonable' solutions considered in [5, 25] would obviously fulfill 
 the criterion of {\em regular} solutions considered in the present paper.},
 are still not known [25], nevertheless, by imposing some (but not all) of the standard `regularity conditions' 
 (which are easily possible), the 
`restricted generating function' can generate all 
 regular\footnote[5]
 {the term `regular' used in refs.[24-26], in fact, denotes positive finite  pressure and density at the origin which do
 not necessarily decrease monotonically outwards (cf. the term `regular' used in the present paper). Consequently, 
 the term `all regular static spherically symmetric perfect fluid solutions' used in refs.[24-26] would, therefore,
  represent both, (i) the `self-bound regular static spherically symmetric perfect fluid solutions' which are discussed in 
 the present paper [category 2(a)], and (ii) the `self-bound static spherically symmetric perfect fluid solutions' with finite
 central densities which are not relevant to the context of the present study (see, footnote `a' of the text).} 
 static spherically symmetric perfect fluid solutions of Einstein's equations [24-26]. 
 Therefore, the type of the algorithm discussed here guarantees a perfect fluid solution, but it does not 
 guarantee a `physically reasonable' perfect fluid solution [24-26].
 For example, Rahman and Visser [24] have investigated the `general quadratic ansatz', which yields the  new form
  of the Goldman I solution [6], and also turns out to be equivalent to Glass-Goldman solution [29] (called G-G in [5]).
  Furthermore, in various regions of parameter space, this general solution reduces to
  at least six other well known exact solutions including Stewart's solution [7], and interior Schwarzschild
 solution. But, in spite of this striking feature of the `general quadratic ansatz', except interior Schwarzschild
 solution (which we have already considered as a basis of our study [3]), both of the generated, insular, perfect fluid solutions 
 (namely, Goldman I and Stewart's solution) are irrelevant to the context of 
 the present study$^{\rm a}$. However, Lake [25] has investigated a particular form of the `generating function'
 which could generate an infinite number of previously unknown ($N > 5$), but physically
 interesting exact solutions (of a particular `class') [including some well known solutions already present in the literature].
 Such an algorithm consider the usual metric (by using geometrized units) in `curvature coordinates' as [25]
 
 \begin{equation} 
 ds^2  = -e^{2\Phi(r)} dt^2  + {[1 - (2m(r)/r)]}^{-1} dr^2  + r^2 d\theta^2 + r^2\sin^2 \theta d\phi^2,
 \end{equation}

where $\Phi(r)$ is also called the (gravitational) `red-shift' parameter, and $m(r)$ is the well known `mass inside radius $r$',
defined earlier. The algorithm for constructing all possible static spherically symmetric perfect fluid solutions of Einstein's 
equations, in fact, connects these two parameters in the following manner [25]
\begin{equation}
m(r) = \frac{\int b(r)e^{\int a(r)dr}dr + C}{e^{\int a(r)dr}},
\end{equation}
where
\begin{equation}
a(r) \equiv \frac{2r^2[\Phi''(r) + \Phi'(r)^2] - 3r\Phi'(r) - 3}{r[r\Phi'(r) + 1]},
\end{equation}
and
\begin{equation}
b(r) \equiv \frac{r[r(\Phi''(r) + \Phi'(r)^2) - \Phi'(r)]}{r\Phi'(r) + 1}.
\end{equation}
 
Here $C$ is a constant, and a prime denotes differentiation with respect to $r$. Eq.(39) is supplemented by
the following boundary conditions [25]:
$\Phi(0) = {\rm finite\,\,\, constant}$ (set by the scale of $t$); $\Phi'(0) = 0$; $\Phi''(0) > 0$; and $\Phi'(r) \neq 0$ for $r > 0$.
That is, the source function $\Phi(r)$ must be a monotone increasing function with a regular minimum at $r = 0$.
In order to have a finite boundary $R$, we further require that $\Phi'(r = R) = M/R(R - 2M)$, where $M = m(r = R)$.

For any monotonically increasing source function $\Phi(r)$ with a regular minima at $r = 0$, Eq.(39) necessarily 
gives a regular$^{\rm d}$ static spherically symmetric perfect fluid solution of Einstein's field equations.
In the present context, the term exact solution is used for those for which Eq.(39) can be evaluated without recourse
to numerical methods. Lake [25] has shown that for the monotonically increasing source function (with a regular
minima at the origin), given by the equation
\begin{equation}
\Phi(r) = \frac{1}{2}N{\rm ln}(1 + \frac{r^2}{\alpha}),
\end{equation}
where $N$ is an integer ($\geq$) and $\alpha$ is a constant ($>0$), can generate an infinite number of previously
unknown ($N > 5$) but physically interesting perfect fluid solutions. The previously known exact solutions 
corresponding to different
values of $N$ are respectively, $N = 1$ (Tolman's type IV solution [9]), $N = 3$ (Heintzmann's solution [30], called Heint IIa in [5]), $N = 4$
(Durgapal IV solution [31], called Durg IV in [5] which is also equivalent to Durgapal and
Bannerji solution [32] called D-B in [5]), and $N = 5$ (Durgapal V solution [31], called Durg V in [5] which is also
equivalent to Durgapal and Fuloria solution [18], called D-F in [5]). Since all these solutions are well
known and also relevant to the
context of the present study (corresponding to a positive finite density at the origin which decreases
monotonically outwards [5]), we will discuss among them, the cases corresponding to the values of $N =1, 4, \,{\rm and}\,\, 5$
respectively.

\subsection{The case N = 1 (Tolman's type IV solution)}
The total mass $M$ for Tolman's type IV solution yields from Eq.(39) (which may be rearranged in terms of compactness ratio
$u(\equiv M/R)$ and the surface density $E_R$ of the solution [33]) in the following form
\begin{equation}
M = u\Bigl [\frac{3u(1 - 2u)}{4\pi E_R(1 - u)} \Bigr ]^{1/2}.
\end{equation}
The relation between surface density $E_R$ and the central density $E_0$ 
in given by the equation
\begin{equation}
\frac{E_R}{E_0} = \frac{2(1 - 2u)(1 - 3u)}{(1 - u)(2 - 3u)}.
\end{equation}
This solution is applicable for the values of $u \leq (1/3)$.
It is evident from Eq.(43) that for an assigned value of the compactness ratio $u$, the mass $M$ can be `measured'
from the knowledge of the surface density $E_R$, which is also dependent upon the central density $E_0$ via Eq.(44). Thus, the mass given 
by Eq.(43) does not represent the `actual mass' appears in the exterior Schwarzschild solution, because the `type
independence' property of the mass (which requires that surface density should be independent of the central density)
 is violated in this case. This is what we have discussed under section 3 of the present study.

\subsection{The case N = 4 (Durg IV = Durgapal and Bannerji (D-B) solution)}
The total mass $M$ for D-B solution yields
in the following form (after rearranging the terms, similarly as in the $N = 1$ case, mentioned above)
\begin{equation}
M = \frac{3X}{16(1 + X)^2}\Bigl [\frac{3X(3 + X)}{\pi E_R} \Bigr ]^{1/2},
\end{equation}
where $X$, in terms of $u$, is given by
\begin{equation}
u = \frac{3X}{4(1 + X)},
\end{equation}
and the surface density $E_R$ is connected with the central density $E_0$ in the following manner
\begin{equation}
\frac{E_R}{E_0} = \frac{3 + X}{3(1 + X)^2}.
\end{equation}
This solution is applicable for the values of $u \leq 0.4214$.
It follows from Eqs.(45) and (46) that for an assigned value of $u$, the total mass $M$ can be worked out from the 
knowledge of the surface
density $E_R$, which is also a function of the central density $E_0$ as indicated by Eq.(47). Evidently, the mass given
by Eq.(45) does not represent the `actual mass' required for the state of hydrostatic equilibrium as explained earlier.

\subsection{The case N = 5 (Durg V = Durgapal and Fuloria (D-F) solution)}
As in the case $N = 1$, and 4 above, the total mass $M$ for D-F solution yields  
in the following form
\begin{equation}
M = \frac{4X(3 + X)}{7(1 + X)^2}\Bigl [\frac{X(9 + 2X + X^2)}{7\pi E_R(1 + X)^3}\Bigr ]^{1/2},
\end{equation}
where $X$ is connected with the compactness ratio $u$ in the following manner
\begin{equation}
u = \frac{4X(3 + X)}{7(1 + X)^2},
\end{equation}
and the relation, connecting surface density $E_R$ and the central density $E_0$ is given by
\begin{equation}
\frac{E_R}{E_0} = \frac{9 + 2X + X^2}{9(1 + X)^3}.
\end{equation}
This solution is applicable for the values of $u \leq 0.4265$. However, 
Eqs.(48) and (49) clearly indicate that for an assigned value of $u$, the total mass $M$ depends upon (and not 
{\em only upon}) the surface
density $E_R$ which is not independent of the central density $E_0$, because of the existence of relation (50). Thus,
the total mass $M$ in this case also, does not represent the `actual mass' which is required by the exterior Schwarzschild
solution.

Let us denote the compactness ratio of homogeneous density distribution by $u_h$, and that of the exact solutions 
corresponding  to  the sub-sections 5.1 - 5.3 by $u_{T-IV}, u_{D-B},$ and $u_{D-F}$ 
respectively. Solving these solutions for various assigned values of the ratio of central pressure to central (energy) density,
$\sigma$, we obtain the corresponding values of the compactness ratio as shown in Table 1. It is seen that for each  and every assigned value of $\sigma$, 
the values corresponding to
$u_{T-IV}, u_{D-B},$ and $u_{D-F}$ respectively, always turn  out to be higher than $u_h$. Or, in other words, the configurations
defined by Tolman's type IV solution (the case $N = 1$), D-B solution (the case $N = 4$), and D-F solution (the case $N = 5$) respectively, do not show 
consistency with the
`compatibility criterion' which is also consistent with the analysis presented under section 3, and its demonstration
carried out under sections 4 and 5 respectively. However, similar results may be obtained for any other permissible
value of $N > 5$.

\begin{table}[h]    
\tbl{ Various values (round off at the fourth decimal place) of the compactness ratio $u(\equiv M/R)$ as
obtained for different assigned values of the ratio of the centre pressure to centre energy-density, $\sigma[\equiv (P_0/E_0)]$,  corresponding to the self-bound regular
solutions, namely - Tolman's type IV [9] solution [indicated by $u_{T-IV}$], Durg IV = Durgapal and Bannerji (D-B) [31] solution [indicated by $u_{D-B}$], and Durg V = Durgapal
and Fuloria (D-F) [18] solution [indicated by $u_{D-F}$] respectively. The compactness ratio corresponding to homogeneous density distribution (interior Schwarzschild solution) is indicated by $u_h$ for the same value of $\sigma$. It is seen that for each and every assigned value of $\sigma$,
 $u_{T-IV}, u_{D-B}$, and $u_{D-F} > u_h$ which is the evidence that the self-bound regular solutions (indicated by $u_{T-IV},
 u_{D-B}$, and $u_{D-F}$ respectively) are not compatible with the state of hydrostatic equilibrium.}
{\begin{tabular}{@{}ccccccccc@{}} \toprule
${\sigma(\equiv P_0 / E_0)}$ && $u_h$   &&  $u_{T-IV}$  && $u_{D-B}$  && $u_{D-F}$ \\ \colrule

0.1252  && 0.1654 && 0.1820  && 0.1743 && 0.1718  \\

0.1859  && 0.2102 && 0.2387  && 0.2221 && 0.2187  \\

0.2202  && 0.2301 && 0.2652  && 0.2429 && 0.2392  \\

0.2800  && 0.2580 && 0.3043  && 0.2714 && 0.2676   \\

0.3150  && 0.2714 && 0.3239  && 0.2847 && 0.2809   \\

(1/3)   && 0.2778 && (1/3)   && 0.2909 && 0.2872   \\

0.3774  && 0.2914 && && 0.3038 && 0.3003  \\

0.4350  && 0.3062 && && 0.3176 && 0.3145  \\

0.4889  && 0.3178 && && 0.3281 && 0.3253  \\

0.5499  && 0.3289 && && 0.3378 && 0.3354  \\

0.6338  && 0.3415 && && 0.3485 && 0.3465  \\

0.6830  && 0.3476 && && 0.3535 && 0.3519  \\

0.7044  && 0.3501 && && 0.3555 && 0.3541   \\

0.7085  && 0.3506 && && 0.3559 && 0.3545  \\

0.7571  && 0.3558 && && 0.3601 && 0.3589  \\

0.8000  && 0.3599 && && 0.3633 && 0.3624   \\

0.8360  && 0.3630 && && 0.3658 && 0.3650   \\ \botrule
\end{tabular}}
\end{table}

\section{Results and Conclusions}

Thus, based upon the analysis regarding the actual mass $`M'$ under section 3, and 
subsequently its demonstration and verification under sections 4 and 5 respectively, we conclude that

\medskip

(i) The self-bound
regular structures corresponding to a single EOS or exact solution can not exist,
because they can not fulfill the requirement of the actual mass $`M'$ set up by exterior Schwarzschild solution.
This finding is consistent
with the results obtained by using the `compatibility criterion' [3]. 

\medskip

(ii) The only
self-bound regular structure which can exist in the state of hydrostatic equilibrium
is described by the homogeneous density throughout the configuration (i.e., the homogeneous
density solution).

\medskip

(iii) The self-bound structures which could exist in the state of hydrostatic equilibrium (this fact 
is also evident from the `compatibility criterion') would always correspond to singularities at the centre (because pressure and density
both become infinity at $r = 0$). 

\medskip

and,

\medskip

(iv) The gravitationally-bound structures naturally fulfill the property of the actual mass $`M'$
required by the exterior Schwarzschild solution, and this finding is also consistent with the
`compatibility criterion' mentioned above.

\section*{Acknowledgments}

The author acknowledges Aryabhatta Research Institute of Observational Sciences (ARIES), Nainital for providing 
library and computer-centre facilities. He wishes to thank the referee for his valuable comments and in particular,
for drawing his attention to the study of refs. [24-25]. S. Joshi, J. C. Pandey, and S. B. Pandey are acknowledged
for their help in arranging some literature related to this work.

\end{document}